\documentclass[12pt,preprint]{aastex}

\def\lesssim{\mathrel{\hbox{\rlap{\hbox{\lower4pt\hbox{$\sim$}}}\hbox{$<$}}}}
\def\gtrsim{\mathrel{\hbox{\rlap{\hbox{\lower4pt\hbox{$\sim$}}}\hbox{$>$}}}}
\providecommand{\etal}{et~al.}

\begin{document}
\title{Multi-Wavelength Properties of the Type IIb SN~2008ax}
\author{P.~W.~A. Roming\altaffilmark{1}, T.~A. Pritchard\altaffilmark{1}, 
P.~J. Brown\altaffilmark{1}, S.~T. Holland\altaffilmark{2,3}, 
S. Immler\altaffilmark{4,5}, C.~J. Stockdale\altaffilmark{6},
K.~W. Weiler\altaffilmark{7}, N. Panagia\altaffilmark{8,9,10}, 
S.~D. Van Dyk\altaffilmark{11}, E.~A. Hoversten\altaffilmark{1}, 
P.~A. Milne\altaffilmark{12}, S.~R. Oates\altaffilmark{13},
B. Russell\altaffilmark{14}, C. Vandrevala\altaffilmark{6}} 

\altaffiltext{1}{Department of Astronomy \& Astrophysics,
Penn State University, 525 Davey Lab, University Park, PA
16802, USA; Corresponding author's e-mail: roming@astro.psu.edu}

\altaffiltext{2}{Universities Space Research Association, 10227 Wincopin Circle, 
Suite 500, Columbia, MD 21044, USA}

\altaffiltext{3}{Center for Research \& Exploration in Space Science \& 
Technology, Code 668.8, Greenbelt, MD, 20771, USA}

\altaffiltext{4}{NASA/Goddard Space Flight Center, Astrophysics Science 
Division, Code 662, Greenbelt, MD 20771, USA}

\altaffiltext{5}{Department of Astronomy, University of Maryland, 
College Park, MD 20742, USA}

\altaffiltext{6}{Physics Department, Marquette University, P.O. Box 1881, 
Milwaukee, WI 53201-1881, USA}

\altaffiltext{7}{Naval Research Laboratory, Code 7210, Washington, 
DC 20375-5320, USA}

\altaffiltext{8}{Space Telescope Science Institute, 3700 San Martin 
Dr, Baltimore MD 21218, USA}

\altaffiltext{9}{INAF-CT Osservatorio Astrofisico di Catania, Via S.
Sofia 79, I-95123 Catania, Italy}

\altaffiltext{10}{Supernova Ltd, OYV \#131, Northsound Rd, Virgin 
Gorda, British Virgin Islands}

\altaffiltext{11}{Spitzer Science Center, IPAC, California Institute 
of Technology, M/C 220-6, Pasadena, CA 91125}

\altaffiltext{12}{Department of Astronomy, University of Arizona, Tucson, AZ
85721, USA}

\altaffiltext{13}{Mullard Space Science Laboratory, University College London, 
Holmbury St. Mary, Dorking, Surrey RH5 6NT, UK}

\altaffiltext{14}{Department of Physics, University of Maryland, 
College Park, MD 20742, USA}

\begin{abstract}
We present the UV, optical, X-ray, and radio properties of the Type IIb
SN~2008ax discovered in NGC~4490. The observations in the UV 
are one of the earliest of a Type
IIb supernova (SN). On approximately day four after the explosion, a dramatic upturn in the 
$u$ and $uvw1$ (${\rm \lambda_c} = 2600 {\rm ~\AA}$) light curves occurred 
after an intitial rapid decline which is attributed to adiabatic cooling
after the initial shock breakout. This rapid decline and upturn is 
reminiscent of the Type IIb SN~1993J on day six after the explosion. 
Optical/near-IR spectra taken around the peak reveal prominent ${\rm H\alpha}$, HeI,
and CaII absorption lines. A fading 
X-ray source is also located at the position of SN~2008ax, implying an 
interaction of the SN shock with the surrounding circumstellar material
and a mass-loss rate of the progenitor of $\dot{M} = (9\pm3) \times 10^{-6} 
{\rm ~M_{\odot}~yr^{-1}}$. The unusual time evolution ($14 {\rm ~days}$) 
of the $6 {\rm ~cm}$
peak radio luminosity provides further evidence that
the mass-loss rate is low.
Combining the UV, optical, X-ray, and radio data with models of helium
exploding stars implies the progenitor
of SN~2008ax was an unmixed star in an interacting-binary.
Modeling of the SN light curve suggests a
kinetic energy ($E_k$) of $0.5\times 10^{51} {\rm ~ergs}$, an ejecta
mass ($M_{ej}$) of $2.9 {\rm ~M_{\odot}}$, and a nickel mass ($M_{Ni}$) 
of $0.06 {\rm ~M_{\odot}}$.  
\end{abstract}


\section {Introduction}
Type IIb supernovae (SNe), first proposed by \citet{WSE87}, were
suggested to be the result of the core-collapse of a Type Ib progenitor 
that has a small, but non-negligible ($\sim 10^{-1}
{\rm ~M_{\odot}}$; Pastorello {\etal} 2008; hereafter
P08), H-envelope. This SN type is arguably one of the 
rarest and most interesting. Only $\sim40$ have been 
discovered in the last 20+ years and only a handful, such as 
SNe~1987K \citep{FAV88}, 1993J \citep{NK93}, 1996cb \citep{QY99}, 
and 2008ax (P08), have been well observed.

In this Letter, we report on imaging observations of 
SN~2008ax with the {\em Swift} \citep{GN04} Ultra-Violet/Optical 
Telescope \citep[UVOT;][]{RPWA05} and X-Ray Telescope 
\citep[XRT;][]{BDN05}, the {\em Chandra} Advanced CCD 
Imaging Spectrometer (ACIS), and the Very Large Array (VLA), 
as well as spectroscopic observations 
with the Hobby-Eberly Telescope \citep[HET;][]{RLW98}. 
The UVOT observations are one 
of the earliest reported observations in the UV of a Type IIb 
SNe, second only to the International Ultraviolet Explorer (IUE) 
observations of SN~1993J \citep{dBKS93}. 


\section{Observations and Data Reductions}
\begin{figure}
\epsscale{0.9}
\plotone{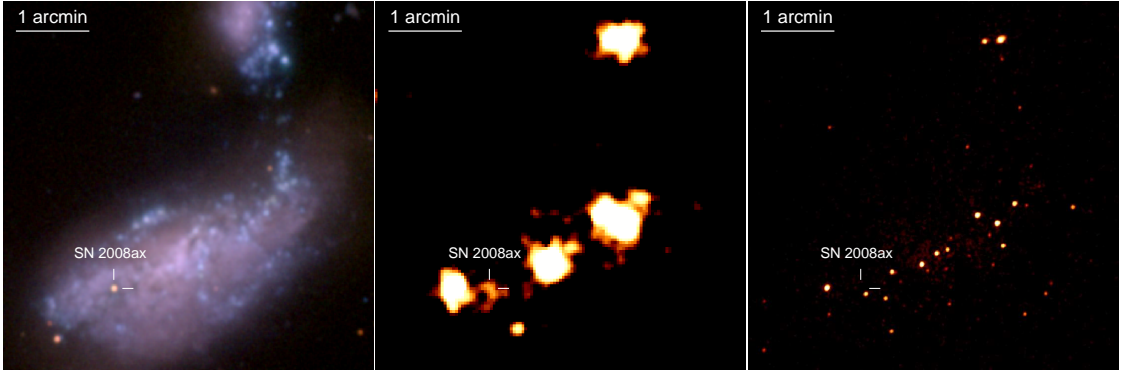}
\caption{{\em Swift} UVOT (left), {\em Swift} XRT (middle), 
and {\em Chandra} (right) images of SN~2008ax and its host 
galaxy. {\em Left:} The UV/optical
image was constructed from the UVOT $b$ ($284 {\rm ~s}$; red), 
$u$ ($285 {\rm ~s}$; green), and 
$uvw1$ ($570 {\rm ~s}$; blue) filters obtained on 
2008-03-04. Positions are 
indicated by white tick marks. {\em Middle:} X-ray image 
($0.2-10 {\rm ~keV}$) from the merged $51.3 {\rm ~ks}$ 
XRT data.
The image is smoothed with a Gaussian filter of 3 pixels ($7\farcs1$) FWHM.
{\em Right:} X-ray image ($0.2-10 {\rm ~keV}$) constructed 
from the $97.6 {\rm ~ks}$ ACIS-S pre-explosion data 
smoothed with a Gaussian filter of 
3 pixels ($1\farcs5$) FWHM.}
\label{fig-2008ax}   
\end{figure}

SN~2008ax (Figure~\ref{fig-2008ax}) was discovered on 2008-03-3.45 
(UT) in NGC 4490 at ${\rm R.A._{J2000}} = 12^{h}30^{m}40^{s}.80$,
${\rm Dec._{J2000}} = +41^{\circ} 38\arcmin 14\farcs5$ \citep{MR08}. 
Based on this detection time and the non-detection of the SN approximately 
six hours earlier, P08 set the time of the shock breakout at ${\rm JD} = 2454528.80\pm
0.15$, which is adopted here.

\begin{figure}
\includegraphics[width=.5\textwidth, angle=90]{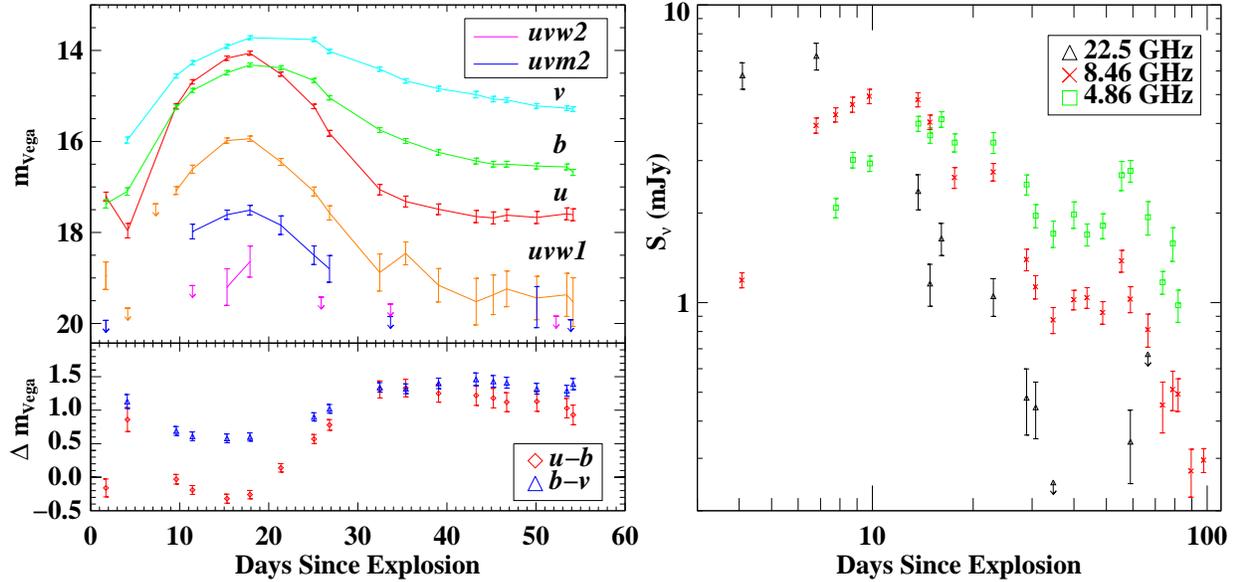}
\caption{{\em Upper-Left:} UV/optical galaxy subtracted 
light curves (not extinction or k-corrected) of SN~2008ax. 
{\em Lower-Left:} $(b-v)_{obs}$ and $(u-b)_{obs}$ color
curves. {\em Right:} Radio light curves. 
Phases are computed since explosion. }
\label{fig-LC}   
\end{figure}

The UVOT observed SN~2008ax from 2008-03-04 (${\rm JD} = 2454530.16$) 
to 2008-04-26 (${\rm JD} = 2454582.52$).
Observations were performed with a cadence varying between
$2-6 {\rm ~days}$ using three optical ($u$, $b$, $v$) and three
UV \citep[$uvw2$, $uvm2$, $uvw1$: ${\rm \lambda_c} = $ 1928, 
2246, $2600 {\rm ~\AA}$, respectively;][]{PTS08} filters. 
A later observation on 2008-11-23 was made after the SN had faded for use as
a galaxy subtraction template. Photometry using a $3\arcsec$ source
aperture, including template galaxy flux subtraction, was performed
following the method outlined in \citet{BPJ09}. The data reduction 
pipeline used the HEASOFT 6.6.3 and {\em Swift} Release 3.3 
analysis tools with UVOT zero-points from \citet{PTS08}. The resulting light 
curves are presented in Table~\ref{tab1} and Figure~\ref{fig-LC}. 

We note that for the majority of SNe analyzed with {\em Swift}
analysis tools, UVOT photometry is well behaved. However, in 
the case of SN~2008ax, the exceptionally bright host galaxy underlying 
the SN leads to small offsets in magnitude due to
coincidence losses \citep[cf.][]{FJLA00} in bright 
extended backgrounds. This results in the reported magnitudes
being $0.23\pm0.07$ ($b$-filter) and $0.19\pm0.06$ ($v$-filter) 
fainter than the P08 magnitudes. 
Although this offset influences the UVOT $u$-band as well, the 
$u$-filter has more throughput in the blue
than ground-based $U$ filters. This is the most probable contributor to the 
$\sim0.6 {\rm ~mag}$ increase in the $u$-band peak as compared
to P08. Since the background galaxy count rate is 
significantly lower in the UV, the UV filters aren't affected
by this undersampling. 

To determine the peak magnitude (${\rm m_{Peak}}$) and time of peak (${\rm t_{Peak}}$) 
in each filter, $10^5$ Monte Carlo simulations fitting a cubic
spline to the data points in each filter were performed. The mean
of the distributions were taken as the peak times and magnitudes, with the
standard deviations as the errors (Table~\ref{tab1}). 
Rise times to peak (${\rm t_{Rise}}$) and absolute magnitudes 
(${\rm M_{Peak}}$) were calculated using the time of shock breakout as
the temporal zero point and the distance modulus ($\mu = 29.92\pm0.29 {\rm ~mag}$), 
respectively, as determined in P08. Errors were calculated
in quadrature (Table~\ref{tab1}). None of the magnitudes include
extinction or k-corrections. 
Since $uvm2$ filter observations only provide
two detections, the upper limits were treated as detections for the 
simulations; therefore, ${\rm t_{Peak}}$ and ${\rm t_{Rise}}$ 
are probably lower than reported. 

\begin{deluxetable}{ccccccc}
\tablecolumns{7}
\tabletypesize{\scriptsize}
\tablecaption{SN~2008ax UV/Optical Characteristics
\label{tab1}}
\tablewidth{0pt}
\tablehead{
  \colhead{Time} &
  \multicolumn{6}{c}{Observed Magnitudes} \\
  \colhead{(JD2454500+)} &
  \colhead{$uvw2$} &
  \colhead{$uvm2$} &
  \colhead{$uvw1$} &
  \colhead{$u$} &
  \colhead{$b$} &
  \colhead{$v$} 
}
\startdata
$30.5\pm0.3$ &    (19.93)     &      ---       & $18.95\pm0.30$ & $17.21\pm0.10$ & $17.61\pm0.13$ &      ---      \\
$32.9\pm0.1$ &      ---       &      ---       &    (19.67)     & $17.96\pm0.16$ & $17.10\pm0.08$ & $15.97\pm0.07$\\
$36.1\pm0.0$ &      ---       &      ---       &    (17.37)     &      ---       &      ---       &      ---      \\
$38.4\pm0.1$ &      ---       &      ---       & $17.08\pm0.09$ & $15.22\pm0.05$ & $15.25\pm0.05$ & $14.56\pm0.05$\\
$40.2\pm0.1$ & $17.98\pm0.16$ &    (19.17)     & $16.61\pm0.09$ & $14.69\pm0.05$ & $14.88\pm0.05$ & $14.27\pm0.05$\\
$44.1\pm0.4$ & $17.61\pm0.10$ & $19.20\pm0.40$ & $15.98\pm0.06$ & $14.17\pm0.05$ & $14.49\pm0.05$ & $13.91\pm0.04$\\
$46.6\pm0.1$ & $17.51\pm0.11$ & $18.64\pm0.34$ & $15.94\pm0.06$ & $14.06\pm0.04$ & $14.32\pm0.04$ & $13.72\pm0.04$\\
$50.1\pm0.0$ & $17.84\pm0.20$ &      ---       & $16.45\pm0.08$ & $14.52\pm0.05$ & $14.38\pm0.04$ &      ---      \\
$53.8\pm0.2$ & $18.50\pm0.20$ &      ---       & $17.10\pm0.10$ & $15.23\pm0.05$ & $14.66\pm0.05$ & $13.76\pm0.04$\\
$54.6\pm1.0$ &      ---       &    (19.42)     &      ---       &      ---       &      ---       &      ---      \\
$55.6\pm0.0$ & $18.80\pm0.29$ &      ---       & $17.57\pm0.16$ & $15.82\pm0.06$ & $15.04\pm0.05$ & $14.02\pm0.04$\\
$61.2\pm0.6$ &      ---       &      ---       & $18.88\pm0.41$ & $17.06\pm0.11$ & $15.75\pm0.05$ & $14.41\pm0.05$\\
$62.4\pm1.8$ &    (19.85)     &    (19.57)     &      ---       &      ---       &      ---       &      ---      \\
$64.1\pm0.1$ &      ---       &      ---       & $18.46\pm0.25$ & $17.32\pm0.12$ & $15.99\pm0.05$ & $14.67\pm0.05$\\
$67.8\pm0.1$ &      ---       &      ---       & $19.16\pm0.37$ & $17.49\pm0.12$ & $16.24\pm0.06$ & $14.84\pm0.05$\\
$72.0\pm0.2$ &      ---       &      ---       & $19.52\pm0.51$ & $17.65\pm0.13$ & $16.43\pm0.06$ & $14.97\pm0.07$\\
$74.0\pm0.1$ &      ---       &      ---       & $19.37\pm0.44$ & $17.68\pm0.13$ & $16.50\pm0.06$ & $15.07\pm0.06$\\
$75.5\pm0.0$ &      ---       &      ---       & $19.24\pm0.39$ & $17.62\pm0.13$ & $16.50\pm0.06$ & $15.09\pm0.05$\\
$78.8\pm0.1$ & $19.64\pm0.45$ &      ---       & $19.44\pm0.48$ & $17.67\pm0.13$ & $16.54\pm0.06$ & $15.22\pm0.05$\\
$81.0\pm2.3$ &      ---       &    (19.84)     &      ---       &      ---       &      ---       &      ---      \\
$82.2\pm0.2$ &      ---       &      ---       & $19.37\pm0.47$ & $17.59\pm0.13$ & $16.56\pm0.06$ & $15.27\pm0.05$\\
$82.6\pm0.7$ &    (19.92)     &      ---       &      ---       &      ---       &      ---       &      ---      \\
$75.5\pm0.4$ &      ---       &      ---       & $19.53\pm0.53$ & $17.61\pm0.13$ & $16.68\pm0.07$ & $15.29\pm0.05$\\
\hline
${\rm m_{Peak}~(mag)}$        &  $17.47\pm0.09$ &  $18.46\pm0.35$ &  $15.90\pm0.05$ &  $14.05\pm0.04$ &  $14.30\pm0.04$ &  $13.56\pm0.06$\\
${\rm M_{Peak}~(mag)}$        & $-12.45\pm0.31$ & $-11.21\pm0.45$ & $-14.02\pm0.29$ & $-15.87\pm0.29$ & $-15.62\pm0.29$ & $-16.36\pm0.30$\\
${\rm t_{Peak}~(JD2454500+)}$ &   $46.3\pm1.4$  &   $49.1\pm5.3$  &   $45.6\pm0.5$  &   $46.1\pm0.4$  &   $47.6\pm0.8$  &   $50.4\pm0.7$ \\
${\rm t_{Rise}~(days)}$       &   $17.5\pm1.4$  &   $20.2\pm5.3$  &   $16.8\pm0.6$  &   $17.2\pm0.4$  &   $18.8\pm0.8$  &   $21.8\pm0.7$ \\
\enddata
\tablenotetext{a}{Magnitudes in parenthesis are upper limit.}
\end{deluxetable}

To calculate ${\rm M_{Peak}}$, a spectrum 
of a similar Type IIb SN \citep[1993J;][]{JDJ94} taken near peak 
was de-reddened and red-shifted to the rest frame for use 
as a template. Milky Way (MW) and host extinction were then computed using 
the $E(B-V)$ value from P08 and applied to our spectral template using 
the \citet{CJA89} MW and \citet{PYC92} SMC laws, respectively. 
Extinction and k-corrections (Table~\ref{tab2}) were 
computed using spectrophotometric methods; 
extinction corrections were computed via the subtraction of synthetic 
magnitudes of an unreddened and reddened template spectrum in the 
observed frame, while k-corrections were computed via the subtraction 
of synthetic magnitudes from the unreddened template spectrum in the 
rest and observed frames.

\begin{deluxetable}{lcccccc}
\tablecolumns{7}
\tabletypesize{\scriptsize}
\tablecaption{Extinction \& k-Correction Factors
\label{tab2}}
\tablewidth{0pt}
\tablehead{
  \colhead{} &
  \colhead{$uvw2$} &
  \colhead{$uvm2$} &
  \colhead{$uvw1$} &
  \colhead{$u$} &
  \colhead{$b$} &
  \colhead{$v$} 
}
\startdata
Extinction (Host + MW) & -2.07 & -2.44 & -1.74 & -1.47 & -1.23 & -0.94\\
k-Correction           & -0.06 & -0.09 & -0.09 & -0.09 & -0.02 & +0.01\\
\enddata
\end{deluxetable}

We analyzed all XRT observations obtained between 
2009-03-05 and 2009-04-26. Due 
to a nearby X-ray source ($7''$ from the SN), X-ray counts 
were extracted within a 2-pixel ($4\farcs8$) radius circular region
centered on the optical position of the SN. The background was extracted 
locally to account for detector and sky background, and unresolved emission 
from the host. An X-ray source is detected ($3.4\sigma$-level) in the merged 51.3~ks XRT 
data (Figure~\ref{fig-2008ax}-{\em Middle}) at the position of the SN with a 
PSF, sampling deadtime, and vignetting-corrected net 
count rate of $(6.4\pm1.9) \times 10^{-4}~{\rm cts~s}^{-1}$. Adopting a thermal 
plasma spectrum \citep[$kT = 10$~keV; cf.][]{FLC02} 
and assuming a Galactic foreground column density with no intrinsic absorption 
\citep[$N_{\rm H} = 1.78 \times 10^{20}~{\rm cm}^{-2}$;][]{DL90} 
gives a 0.2--10 keV flux and luminosity of 
$f_{0.2-10} = (3.0\pm0.9) \times 10^{-14}~{\rm erg~cm}^{-2}~{\rm s}^{-1}$ and 
$L_{0.2-10} = (3.3\pm1.0) \times 10^{38}~{\rm erg~s}^{-1}$, respectively
(assuming 9.6~Mpc; P08). Rebinning the data into two 
epochs (2008-03-05 to 2008-03-31 and 2008-04-04 to 2008-04-26) with similar 
exposure times (24.3~ks and 27~ks, respectively) shows the X-ray source 
faded by $\approx 4$ from $L_{0.2-10} = (6.0\pm1.9) \times 10^{38}
~{\rm erg~s}^{-1}$ to $(1.4\pm0.9) \times 10^{38}~{\rm erg~s}^{-1}$ 
during the observations.

We further analyzed archival pre-SN ACIS data obtained on 
2000-11-03, 2004-07-29, and 2004-11-20 to characterize the X-ray contamination of 
the SN with the nearby X-ray source. No 
X-ray source is visible at the position of the SN 
(Figure~\ref{fig-2008ax}-{\em Right}), but the nearby X-ray source 
is clearly detected, with an average luminosity of $L_{0.2-10} = (5.9\pm0.4) 
\times 10^{38}~{\rm erg~s}^{-1}$. Comparison of the X-ray luminosity for each 
of the three epochs shows the source isn't variable. We therefore 
extracted the counts from the position of the SN from the XRT data using a 
larger aperture corresponding to the 100\% encircled energy radius of the XRT 
PSF, which contains the nearby X-ray source, and subtracted the luminosity of 
the {\em Chandra} source. Within the errors, the same residual X-ray luminosity 
for the source at the position of the SN is obtained when compared to the above 
analysis using a smaller aperture.

Radio observations were made with the VLA\footnote{The VLA of the National Radio
Astronomy Observatory is operated by Associated Universities, Inc. under
a cooperative agreement with the National Science Foundation.} at 1.3,
3.6, 6.0, and 20~cm in continuum mode. Initial data reduction and
analysis of the radio data was similar to the methods used for SNe~1993J and 2001gd
\citep{WKW07,SCJ07}.

\begin{figure}
\epsscale{0.9}
\plotone{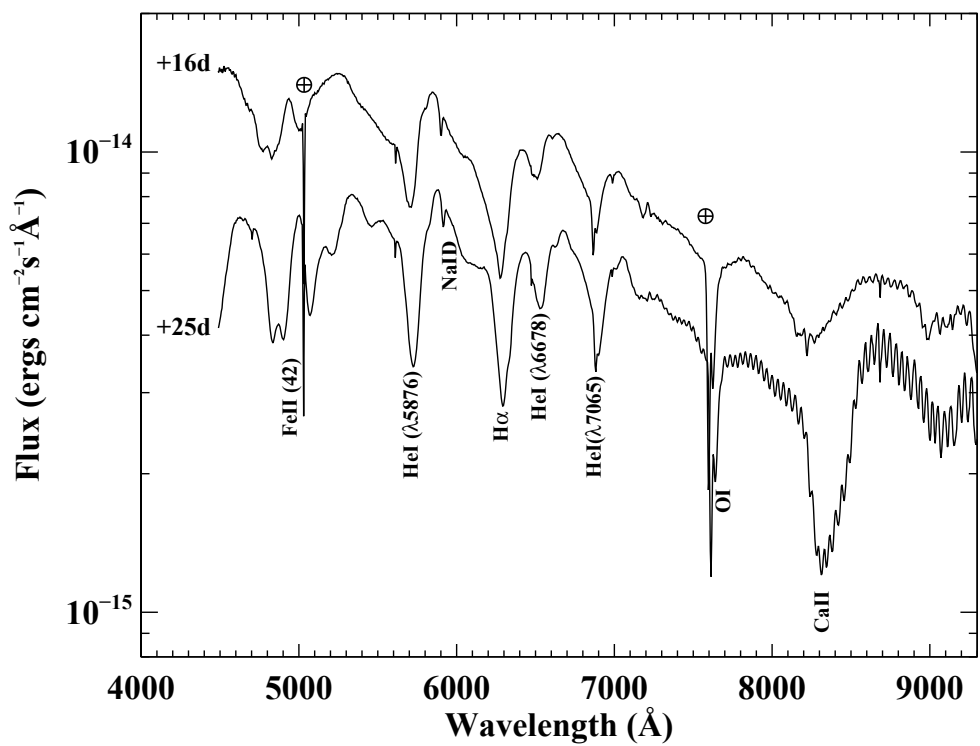}
\caption{HET LRS spectra of SN~2008ax taken 16 and 25 days after explosion. 
Prominent lines specified in \citet{JDJ94} and P08 are indicated. $\bigoplus$ represents
telluric absorption lines. The spectra haven't been corrected for redshift or 
extinction. The +25d spectrum is shifted down by a constant factor of $50\%$ 
for clarity.}
\label{fig-HET}   
\end{figure}

SN~2008ax was observed with the HET 
on 2008-03-19 and 2008-03-28, for 600 and $1200 {\rm ~s}$, respectively.
The Low Resolution Spectrograph \citep[LRS;][]{HGJ98} was used with a 
2\arcsec\, slit (${\rm R}\sim300$; $\Delta\lambda \sim4,500-10,000 {\rm ~\AA}$).
Standard IRAF 
reduction techniques of bias subtraction, flat fielding,
and wavelength calibration were used.
Relative flux calibration was performed using several flux standards
\citep[BD262606, HD84937, \& HZ44;][]{FM96,MG90}
observed during March 2008. The spectra are displayed in Figure~\ref{fig-HET}.

\section{Results}
Observations of SN~2008ax began $\sim1.5 {\rm ~days}$ after explosion
and continued until day 54. The SN was detected in all UVOT color
filters, but after template subtraction, fewer detections remained
in the $uvm2$ and $uvw2$ filters. Assuming a SN~1993J-like UV
spectrum, this is expected since the UV flux is 
intrinsically fainter due to line blanketing, and $E(B-V) = 0.3$
(P08) therefore the UV flux is more suppressed (see Table~\ref{tab2}).

The general trend for the light curves is that the UV bands tend to 
rise more rapidly, peak earlier, and decay more rapidly than the 
optical bands (see Table~\ref{tab1}). This trend is also 
seen in a sample of UV/optical
light curves of Type II SNe \citep{BPJ09}. The $v$ and $b$ SN~2008ax light
curve shapes are consistent with those of P08. A small shoulder
in the $b$ is also seen in the first observation.

Also evident is an initial fading in the $u$ and $uvw1$ light curves
followed by a rise reminiscent of the dip seen in SNe~1987A
\citep[Type II;][]{HM88}, 1993J \citep[IIb;][]{SBP93},
1999ex (Ibc; Stritzinger {\etal} 2002; hereafter
S02), 2008D \citep[Ibc;][]{SAM08}, and 2006aj 
\citep[Ic;][]{CS06}. The same shoulder in the $B$ and the upturn
in the $U$ seen in SN~1999ex on day four after the explosion
(S02) is also seen in the UVOT $b$ and $u$ data for SN~2008ax
on approximately day four. Additionally, a UV upturn is present which is the first
time this has been seen in a Type IIb SN and only the third time
in any SNe \citep[cf. SNe 1987A \& 2006aj;][]{PC95,CS06}. This dip in the light
curve is attributed to adiabatic cooling after the initial
shock breakout (S02), and 
reinforces the idea that the progenitor is the result of a core-collapse of
a massive star (S02). Using HST data, \citet{CRM08}
further constrained that the SN~2008ax progenitor was either a single massive star
that lost most of its H-envelope prior to exploding
or an interacting binary in a low-mass cluster.

Using the method described by \citet{RD06}, we performed a 
Monte Carlo simulation with $2.5 \times 10^5$ realizations of the SN~2008ax $v$-band light 
curve\footnote{The \citet{JDJ99} portion of the model used by 
\citet{RD06} was excluded from our simulation since no late time data 
are included in our sample. The \citet{AWD82} portion of the
model alone fit the light curves reasonably well.} by varying three
model parameters: kinetic energy ($E_k$), ejected mass ($M_{ej}$), 
and ejected nickel mass ($M_{Ni}$). The
best $\chi^2$ fit resulted in $E_k = 0.5\times10^{51} {\rm ~ergs}$, 
$M_{ej} = 2.9 {\rm ~M_{\odot}}$, and $M_{Ni} = 0.06 {\rm ~M_{\odot}}$. 
The resultant parameters are consistent with values calculated by \citet{RD06} for 
other Type IIb SNe, except that the SN~2008ax value for $M_{ej}$ is
$\sim3\times$ higher.

A ($u-b$) and ($b-v$) color plot (Figure~\ref{fig-LC}-{\em Lower-Left})
tracks temperature variations in the photosphere 
(Stritzinger {\etal} 2009; hereafter S09). 
Initially, the blue to red jump ($\Delta(u-b) = 1.0$) is due to the
adiabatic cooling discussed previously. This is followed by a transition
back to the blue as the photospheric temperature increases and the 
light curves reach maximum. As the temperature decreases, the 
color curves transition back to the red and begin to 
level off. Of note is that the $u$ and $uvw1$ light curves stop 
declining in brightness at $\sim25 {\rm ~days}$ 
after maximum. This corresponds with the color evolution to the
blue in the $(u-b)$ at the same time. This blue excess was also
seen in SNe 2007Y and 2008aq, and was attributed to a
change in the opacity of the ejecta, or shock heating produced
by interaction of high-velocity SN ejecta with circumstellar material 
(CSM; S09). 

The lack of an X-ray source in pre-SN 
{\em Chandra} images and the detection of a fading X-ray source 
in {\em Swift} XRT data confirms that X-ray 
emission is detected from SN~2008ax. These 
X-rays likely arise from interaction of the outgoing SN shock 
with substantial amounts of CSM. If the CSM 
was deposited by a strong stellar wind prior to outburst, 
as expected for the massive progenitor of a SN IIb, the X-ray 
luminosity can be used to estimate the mass-loss rate. Following the 
discussion in \citet{SI07}, 
we infer a mass-loss rate of the progenitor of 
$\dot{M} = (9\pm3) \times 10^{-6} {\rm ~M_{\odot}~{\rm yr}^{-1}}
~(v_{\rm w}/10~{\rm km~s}^{-1})$ for an assumed speed of the outgoing 
shock of $v_{\rm s} = 10,000~{\rm km~s}^{-1}$ and scaled for a stellar wind 
speed of $v_{\rm w}=10~{\rm km~s}^{-1}$. 
Only two other SN IIb have been detected in X-rays over the past four 
decades: SNe~1993J \citep{CP08} and 2001gd \citep{PT05}. In 
contrast, SN~2008ax is only a weak X-ray emitter, one and two orders 
of magnitude fainter than SNe~2001gd and 1993J, respectively. 
Subsequently, the inferred mass-loss rate for the 
progenitor of SN~2008ax is significantly below that of SN~1993J 
\citep[$10^{-5} - 10^{-4} {\rm ~M_{\odot}}~{\rm yr}^{-1}$;][]{SI01} 
and more characteristic of SNe IIP \citep[$10^{-6} - 10^{-5} 
{\rm ~M_{\odot}}~{\rm yr}^{-1}$; cf.][]{CFN06,SI07}. This wide 
range of mass-loss rates for SNe 
IIb could be caused by a diversity of binary progenitor 
systems\footnote{Note: SN 1993J doesn't show signs in its radio evolution of having 
been in a binary system \citep{WKW07}, although other wavebands indicate
otherwise \citep{VDSD04}.}.

The general nature of the SN~2008ax radio light curve 
(Figure~\ref{fig-LC}-{\em Right}) is very typical of 
core-collapse SN \citep[CCSNe;][]{MVI09,WKW02}. However, the time evolution of the 
radio emission is somewhat unusual for a Type IIb radio SN when compared
to SNe~1993J, 2001gd, and 2001ig \citep{WKW07,SCJ07,RSD04,RSD06}. SN~2008ax
reached its 6~cm peak radio luminosity in 14~days, while SN~1993J took 
133~days after explosion to achieve peak \citep{WKW07}.
The exact time to reach this point is uncertain for both SNe~2001gd
and 2001ig, but the parameterized models outlined in \citet{WKW02} indicate
the peak occurred at $\sim100 {\rm ~days}$ \citep{SCJ07,RSD04}. In contrast,
the first event associated with a Type Ic SN 
and a gamma-ray burst, SN 1998bw/GRB 980425, reached its peak 6~cm 
radio luminosity in 10 days \citep{WKW01}.

A typical radio-derived mass-loss rate for the earlier Type IIb SNe is 
$10^{-5} {\rm ~M_{\odot}~{\rm yr}^{-1}}$
\citep{WKW02}. The exact radio-derived mass-loss rate for SN~2008ax is difficult to
establish as the exact nature of the SN progenitor is ambiguous \citep{CRM08}.  
Following the 
discussion in \citet{WKW02}, 
we infer a range of possible mass-loss rate of the progenitor of 
$\dot{M} = (1-6) \times 10^{-6} {\rm ~M_{\odot}~{\rm yr}^{-1}}
~(v_{\rm w}/10~{\rm km~s}^{-1})$ for an assumed speed of the outgoing 
shock of $v_{\rm s} = 10,000~{\rm km~s}^{-1}$ and scaled for a stellar wind 
speed of $v_{\rm w}=10~{\rm km~s}^{-1}$. The ambiguity of the nature of the progenitor makes the 
wind speed of the progenitor-established CSM
very uncertain. 

Furthermore, the radio evolution of SN~2008ax 
appears to show short time scale modulations of $5-10 {\rm ~days}$
(see Figure~\ref{fig-LC}-{\em Right}).
This may be similar to the evolution of SN~2001ig whose progenitor
was attributed to an interacting-binary system 
or the evolution of SN 1998bw whose progenitor was a very massive star \citep{RSD04,RSD06,WKW01}. 
A deeper analysis of the radio evolution of SN~2008ax is on-going.

Spectroscopic observations of SN~2008ax occurred 16 and 25 days after explosion.
The HET spectra (Figure~\ref{fig-HET}) are very similar --- 
with prominent ${\rm H}\alpha$ and HeI lines --- to SNe~1993J
\citep{JDJ94} and 2008ax (P08) spectra taken 18 and 28 days
post explosion, respectively. 
P08 showed at 56 days that CaII was also present in SN~2008ax,
but presented no data redward of $\sim8000 {\rm ~\AA}$ previous to this
time. HET spectra reveal that the CaII absorption line was 
significant at 16 days ($\sim3$ days before $b$-peak)
and was a factor of $\sim3$ larger nine days later. This
behavior is reminiscent of SNe~2005bf \citep[Ibc;][]{FG06} and 
2007Y (Ib; S09). Both SNe had very early strong
CaII absorption lines, followed by a decrease in strength, and
then a rapid increase in strength. Assuming the early time
CaII absoprtion in SN~2008ax was also strong and then weakened,
the HET spectra indicate around peak and after peak 
that CaII is caused by a photospheric,
rather than a high velocity component above the photosphere 
\citep[S09;][]{PJ07}.
The behavior of the blueward components of the HET spectra,
particularly the strengthening of the HeI lines, are 
consistent with the P08 results.

\section{Conclusion}
The $u$ and $uvw1$ light curves of SN~2008ax show a rapid initial
decline to a minimum approximately four days after shock breakout.
The SN then brightens to maximum light in the $u$-band at $17.2 \pm
0.4$ days after shock breakout. This behaviour is seen in other
CCSNe and is predicted in numerical models of CCSNe with stripped atmospheres
\citep{SNT1990,SSK1994}. However, it isn't seen in
most CCSNe light curves. This may be due to the paucity of
detailed early-time observations for most SNe, or indicate a
physical difference in the progenitors. The dip is attributed to 
rapid adiabatic cooling of the photosphere, which is accelerated
by the shock passage. The time scale of adiabatic cooling
depends on the volume of the photospheric shell, and thus the
radius of the progenitor. Modelling of CCSNe with stripped
atmospheres suggests that the adiabatic cooling acts on time scales of
several days, which is in agreement with the UV minimum seen in 
SN~2008ax.

\citet{SNT1990} produced models of exploding helium stars, which are
believed to produce SNe IIb. The model assumes the progenitor is a
stripped star in a binary system. From the X-ray and radio mass-loss 
rates coupled with the short time evolution of the radio
emission, the SN~2008ax progenitor appears to have been in an 
interacting binary, consistent with the work of \citet{CRM08}.
The observed time of maximum light 
and shape of the later SN~2008ax light curve are consistent with an
unmixed progenitor model which produces early UV dips lasting 
longer than those with mixing in the
progenitor. Dips in mixed models tend to have minima at $\lesssim
5$ days while those in unmixed models tend to have minima at
$\approx$ 5--10 days.  The time of the SN~2008ax dip
is consistent with a mixed progenitor; however, the time is 
uncertain due to the lack of very early time
data. It's therefore possible that the true minimum occurred
later than $\approx 4$ days, as is predicted for unmixed models.
Further evidence for an unmixed progenitor comes from the strength of
the dip. The observed dip is $\approx 4$ magnitudes consistent 
with an unmixed model. These
three lines of evidence suggest that the progenitor of SN~2008ax was
either unmixed, or very lightly mixed.

\citet{SNT1990} find that helium
stars with less mixing tend to be more massive than those that undergo
extensive mixing. If the progenitor of SN~2008ax is unmixed
then it may be on the more massive side of the progenitor mass
distribution. This supports our result that the mass-loss rate from
SN~2008ax is lower than for a typical Type IIb SNe.

The onset and strength of the adiabatic cooling dip increases as
the progenitor mass increases, so more massive progenitors have more
prominent adiabatic cooling following the shock breakout. This
suggests that the lack of an observed cooling dip for many SNe IIb may
be due to less massive progenitors. This, combined with the
paucity of high-cadence early-time observations, may explain why the
cooling dip is not seen in many SNe IIb.

\acknowledgments
This work is sponsored at PSU by NASA contract 
NAS5-00136 and at Marquette by NASA award NNX09AC90G. 
The HET is a joint project of UT-Austin, PSU, Stanford, 
Ludwig-Maximilians-Universit\"at M\"unchen, and Georg-August-Universit\"at 
G\"ottingen, and is named in honor of its principal benefactors, 
William P. Hobby and Robert E. Eberly. 


\end{document}